\documentclass[aps,prl,twocolumn,superscriptaddress,groupedaddress]{revtex4}  
\usepackage{graphicx}  
\usepackage{dcolumn}   
\usepackage{bm}        
\usepackage{amssymb}   
\usepackage{slashed}
\usepackage{multirow}
\usepackage{color}

\usepackage{soul}

\begin{document}

\title{Discovering heavy new physics in boosted $Z$ channels: $Z \rightarrow  l ^+l^- $ vs $ Z \rightarrow \nu\bar{\nu}$}

\author{Mihailo Backovi\'{c}}
\affiliation{Center for Cosmology, Particle Physics and Phenomenology - CP3, \\ Universite Catholique de Louvain, 2 Chemin du Cyclotron, Louvain-la-Neuve, Belgium}
\author{Thomas Flacke}
\affiliation{Department of Physics, Korea Advanced Institute of Science and Technology, \\
335 Gwahak-ro, Yuseong-gu, Daejeon 305-701, Korea}
\author{Jeong Han Kim}
\affiliation{Department of Physics, Korea Advanced Institute of Science and Technology, \\
335 Gwahak-ro, Yuseong-gu, Daejeon 305-701, Korea}
\affiliation{IBS Center for Theoretical Physics of the Universe, \\
70, Yuseong-daero 1689-gil, Yuseong-gu, Daejeon, Korea}
\author{Seung J. Lee}
\affiliation{Department of Physics, Korea Advanced Institute of Science and Technology, \\
335 Gwahak-ro, Yuseong-gu, Daejeon 305-701, Korea}
\affiliation{School of Physics, Korea Institute for Advanced Study, \\
 85 Hoegiro, Dongdaemun-gu, Seoul 130-722, Korea}


\date{\today}

\begin{abstract}
\noindent

We propose a strategy for new physics searches in channels which contain a boosted $Z$ boson and a boosted massive jet in the final state. Our proposal exploits the previously overlooked advantages of boosted $Z\rightarrow \nu \bar{\nu}$ topologies, where collimated neutrinos result in signals with large missing energy. We illustrate the advantage of this channel in a case study of singly produced TeV scale charge 2/3 fermionic top partners ($T'$) which decay to $tZ$ final states. A comparison with the di-leptonic channel reveals that, despite the large $t\bar{t}$ background, signals with missing energy combined with jet substructure techniques offer superior probes of new physics at TeV scales. The effect can be attributed to a factor of $\sim 3$ enhancement in the signal cross section, coming from the branching ratio of $Z\rightarrow \nu \bar{\nu}$. We exploit the unique event topology of singly produced top partners to suppress the $t\bar{t}$ background, as well as further improve on the existing proposals to detect $T'$ in the boosted di-lepton channel. 
Our conclusions on advantages of $Z\rightarrow \nu \bar{\nu}$ can be extended to most resonance searches which utilize a boosted $Z$ boson in the final state. 

\end{abstract}

\maketitle
\begin{center}
\textbf{ I. INTRODUCTION}
\end{center}

New physics searches in channels containing a $Z$ boson in the final state are an important part of the LHC physics program, with $WZ$  and $hZ$ productions being some of the important probes of the Standard Model (SM). In the context of new physics searches,  $Z$ boson production accompanied by a $h, \,W$ or top quark are also important. Such final states appear (for example)
in the decays of charge 2/3 vectorlike top quark partners ($T'$), which emerge as generic features in many models that address the hierarchy problem within the framework of Naturalness~\cite{Schmaltz:2005ky,Contino:2003ve,Agashe:2004rs,Agashe:2006at}.   

At the dawn of LHC Run II, a natural question to ask is which of the $Z$ decay modes
will be the most sensitive to new physics at the TeV scale. Conventional wisdom tells us that leptons offer clean signals with low backgrounds. For instance, $T'$ decaying into $t Z$ is a primary option for most experimental searches, where $Z$ decaying into di-leptons has so far been the most sensitive channel  \cite{Aad:2014efa}. Recently, Ref. \cite{Reuter:2014iya} also proposed a search strategy for singly produced $T'$ in a boosted di-lepton channel for Run II of the LHC. While leptons are convenient final states, branching ratios of heavy SM states to leptons are  small, and there is always a lower limit on the signal production cross section (and hence an upper limit on the mass scale) which can be probed by leptons at fixed integrated luminosity. 

In the initial stages of LHC Run II, it hence might be desirable to look for alternative search channels with larger signal cross section and reasonably small backgrounds, in order to improve the prospects of an early discovery. 
In this paper, we will show that  $Z$ decaying to $\slashed{E}_T$ ($Z_{\mathrm{inv}}$) accompanied by a boosted massive jet satisfies this criterion, when the mass scale of new physics is above 1 TeV. 

The $Z_{\mathrm{inv}}$ channel was not used in LHC Run I analyses, as exploration of the mass scales of $O(100)$ GeV focused on $Z$ events with low boost. The large angles between the neutrinos hence resulted in missing energy signals which were too low to be efficiently used for background discrimination. However, with a large enough boost of the $Z$, the $Z_{\mathrm{inv}}$ channel becomes relevant for Run II. The utility of missing transverse energy has been discussed for the $VH$ production in a boosted regime~\cite{Butterworth:2008iy} as well as the searches for Kaluza-Klein gravitons~\cite{Chen:2014oha} in the $ZZ \rightarrow \slashed{E}_T l^+ l^-$ channel. Here, we consider the $Z t$ channel where the top decays hadronically \footnote{This channel is more challenging because it has no hard, isolated leptons in the final state and a potentially $t\bar{t}$ large background.} and systematically compare the leptonic and the invisible $Z$ decay modes.

The core of our proposal is an observation of several important qualitative changes in the phenomenology of LHC Run II.
 
First, for mass scales $ \gtrsim 1 $ TeV, the SM decay products of heavy new particles become highly boosted.  It follows that boosted $Z$s expected at Run II will decay into collimated  neutrinos and hence large missing energy signatures (i.e., in resonance searches one could expect  signatures of $\slashed{E}_T \sim M_{\mathrm{res}} / 2$). 

Second, standard jet reconstruction techniques and lepton isolations will not be adequate and tools of jet substructure and alternative lepton isolation variables will have to be employed. 

Third, the boosted regime will be characterized by different  efficiencies for reconstruction and tagging of the $t,Z,W,h$ decay channels. As a result, experimental sensitivity in different decay modes to new physics searches will change compared to Run I. At the same time, SM backgrounds for signatures of very highly boosted objects (e.g. very high-$p_T$ jets, leptons, di-leptons, large $\slashed{E}_T$, etc.) fall much faster than the signal, implying an altered background rejection power for the different channels.

\begin{figure}[t]
\includegraphics[scale=0.2]{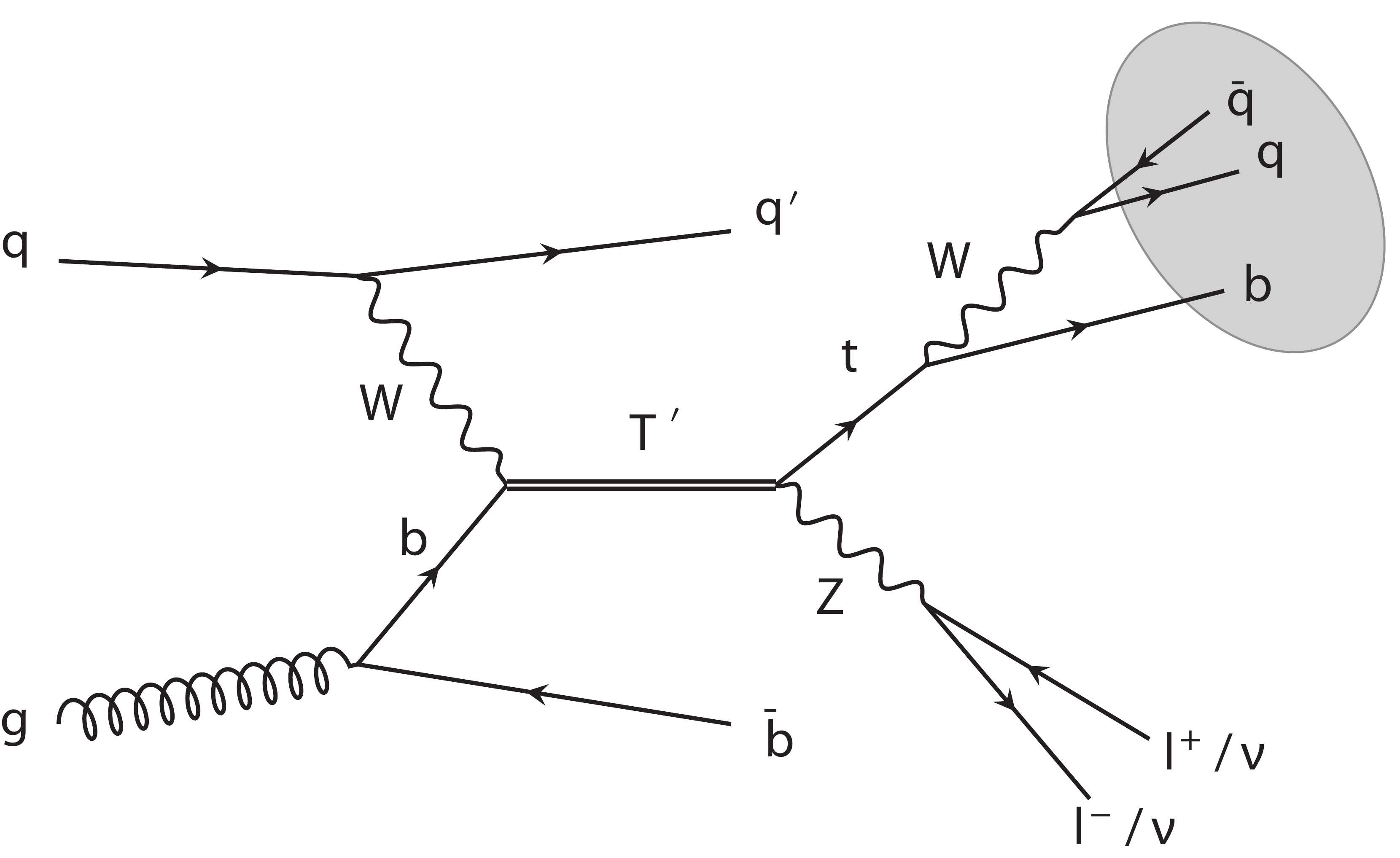}
\caption{Single production channel for a $T'$ decaying into $t Z$.}
\label{fig:prod}
\end{figure}

$T'$ models are an excellent example of studies where the boosted $Z$ regime will be relevant in the future LHC runs. Past studies of ATLAS \cite{ATLASnotes1} and CMS \cite{Chatrchyan:2013uxa} established bounds on mass of the vectorlike top partners, excluding states with mass lighter than $\sim700-800$ GeV (with the precise bound depending on the $T'$ branching fractions). Run II of the LHC will thus probe the TeV mass range, where the boosted regime will become important. For concreteness, here we consider $T'$ single production with subsequent $T'\rightarrow t Z$ decay for which the process shown in Fig.\ref{fig:prod} yields the dominant contribution. 

In the following, we show that considerations of $\slashed{E}_T$ signals greatly extend the ability of Run II of the LHC to discover possible new physics in the boosted $Z$ channel. Although we use $T'$ searches for the purpose of illustration, our main conclusion that at sufficiently high boost $Z_{\mathrm{inv}}$ searches will outperform the  $Z_{ll}$ searches is valid in a more general sense and qualitatively applies to many physics searches which utilize a boosted $Z$ boson in the final state.
\bigskip

\begin{center}
\textbf{II. SAMPLE MODEL AND EVENT SIMULATIONS}
\end{center}
 
For the purpose of event simulation, we use the minimal composite Higgs model with a partially composite top (cf. Ref.\cite{Backovic:2014uma} for the model Lagrangian, parameter definitions and the detailed derivation of the interactions.)  In the singlet-partner-limit, the model contains only one light vectorlike  top partner: an $SU(2)_L$ singlet with charge $2/3$. The top-partner sector of the model is described by the effective Lagrangian\\

\begin{eqnarray}
\mathcal{L} &\supset& \bar{\tilde{T}}\left(i \slashed{D}- M_1\right)\tilde{T}+\bar{q}_{L} i \slashed{D}q_{L}+\bar{t}_{R} i \slashed{D}t_{R} \label{eq:Leff}\\
& - & \left(\lambda_R f \cos (\bar{h} /f) \bar{t}_R\tilde{T}_L -  \frac{\lambda_L f \sin (\bar{h} /f)}{\sqrt{2}} \bar{t}_L\tilde{T}_R+\mbox{H.c.}\right)\, ,\nonumber
\end{eqnarray}
where $\bar{h}= v+h$, $f$ is the Higgs compositeness scale, $M_1$ is the single mass scale and $\tilde{T}$ denotes the gauge eigenstate of the top partner while $T'$ denotes the mass eigenstate.

We consider only $T'$ production, which dominates over $T'$ pair production for high $M_{T'}$ due to larger phase space. For exactly what mass single production becomes dominant is model dependent, but in many models this occurs around $M_{T'}\simeq 1$ TeV for natural parameter choices ({\it i.e.} only slightly above the scale up to which Run I is sensitive).

\begin{figure}[t]
\hspace{-30pt}
\includegraphics[scale=0.45]{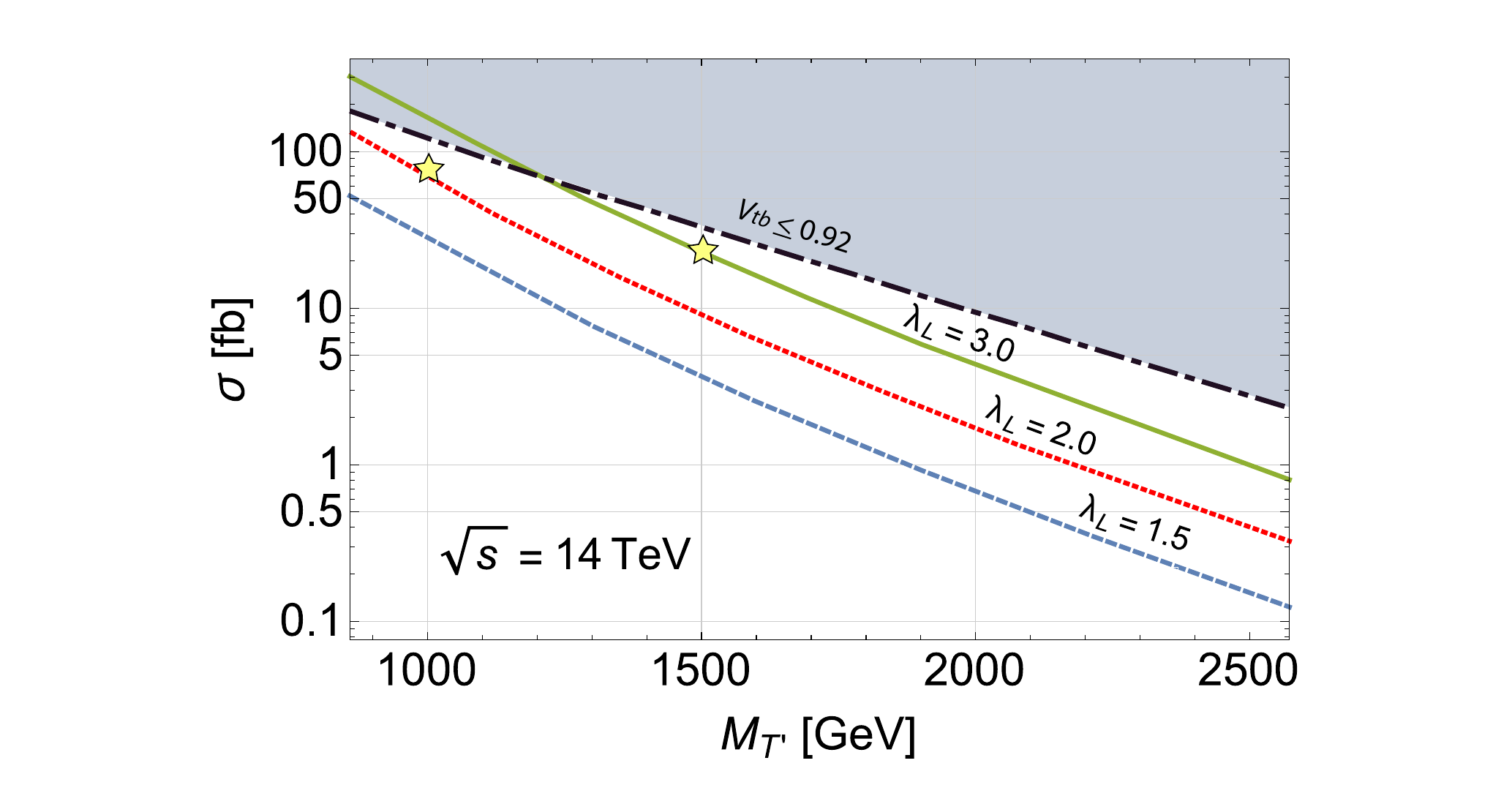}
\caption{Single production cross section for  $T'$ or its conjugate decaying into $t Z$ (or $\bar{t} Z$) in the sample model described by Eq.~(\ref{eq:Leff}), for various values of $\lambda_L$ with $f = 780 $ GeV.}
\label{fig:Xsection}
\end{figure}	

The results of our analysis depend on $M_{T'}$ and $\sigma_{T'}\equiv\sigma( p p \rightarrow T'/\bar{T}' + X)\times BR(T'\rightarrow tZ)$, the production cross section of the $T'$ times the  branching fraction of $T'\rightarrow t Z$ (provided that the width of $T'$ is small), while the dependence of the event kinematics on specific model parameters is small enough to be neglected. For illustration and completeness, Fig.~\ref{fig:Xsection} shows $\sigma_{T'}$ in our sample model as a function of $M_{T'}$ for several values of $\lambda_L$. The two points marked by a star are sample points which we use in our later analysis \footnote{For these plots, we fixed $f= 780$~GeV. $\lambda_R$ is determined by the requirement to obtain the correct top mass ({\it cf. e.g.} Ref.~\cite{Backovic:2014uma}).}.  The shaded region in Fig.~\ref{fig:Xsection} is excluded by the direct bound $|V_{tb}| > 0.92$ obtained from single-top searches \cite{Chatrchyan:2012ep}.\footnote{Stronger indirect bounds on $V_{tb}$ are obtained from flavor physics, but those can be circumvented in composite Higgs models with additional flavor structure in the model ({\it c.f. e.g.} \cite{Cacciapaglia:2015dsa}).} This latter bound on the production cross section applies to our specific sample model only, while it can be relaxed in composite Higgs models with different top partner representations or other vectorlike  quark models with two or more partner multiplets. In the following, we present results in terms of $M_{T'}$ and $\sigma_{T'}$  rather than the input parameters $M_1, \lambda_L, f$ in order to maintain a minimum of model dependence.
 
We simulate signal and backgrounds with leading order \verb|MADGRAPH 5| \cite{Maltoni:2002qb} (using NNPDF2.3LO1 PDFs \cite{Ball:2013gsa}  interfaced with \verb|PYTHIA 6| \cite{Sjostrand:2006za} for parton showering and hadronization, while we conservatively assume a $k$-factor of 2 for all background channels. We match the background samples to extra jets using the 5 flavor so-called MLM~\cite{Mangano:2006rw} matching scheme. 

\begin{center}
\begin{table*}[t]
\begin{tabular}{|c||cccc|cc||cccc|cc|}
\hline
\multirow{2}{*}{$Z\rightarrow \nu \bar{\nu}$}&\multicolumn{6}{c||}{$M_{T'}=1.0$ TeV search}&\multicolumn{6}{c|}{$M_{T'}=1.5$ TeV search}\\ \cline{2-13}
&Signal& \hspace{5pt}$t\bar{t} \hspace{5pt}$&$Z+X$&$Z+t$&$S/B$&$S/\sqrt{B}\, (100 \, \mathrm{fb}^{-1})$& Signal &\hspace{5pt}$t\bar{t}$ \hspace{5pt}&$Z+X$&$Z+t$&$S/B$&$S/\sqrt{B} \,(100 \, \mathrm{fb}^{-1})$\\\hline
Basic cuts                       &3.5                    & 900                  &6100      &11        &0.00050      &0.42              &1.0               &140        &1200      &2.4       &0.00074       &0.27\\
$Ov^t_{3} > 0.6$            &2.7                    &510                   &840        &6.5        &0.0020      &0.75               &0.87             &81         &230         &1.6       &0.0028   &0.49 \\
$b$-tag                          &2.0                    &320                    &16          &4.3        &0.0057      &1.1               &0.54              &45         &3.2        &0.94      &0.011     &0.77   \\
$\slashed{E}_T$-cut 	    &\bf{1.3}              &13                     &5.3         &0.89       &\bf{0.065} &\bf{2.9}            &\bf{0.41}     &1.00      &0.78       &0.14       &\bf{0.21} &\bf{3.0} \\\hline
$N_{\mathrm{fwd}}\geq 1$             &\bf{0.79}     &2.6           &0.74        &0.27      &\bf{0.22} &\bf{4.1}           &\bf{0.28}     &0.20     &0.11        &0.041     &\bf{0.80} &\bf{4.7}\\\hline
$\Delta \phi_{\slashed{E}_T,j}>1.0$&\bf{0.66}  &0.94          &0.58        &0.22       &\bf{0.38}     &\bf{5.0}       &\bf{0.22}     &0.076     &0.083     &0.033    &\bf{1.2}  &\bf{5.1}\\\hline
\end{tabular}

\begin{tabular}{|c||ccc|cc||ccc|cc|}
\hline
\multirow{2}{*}{$Z\rightarrow l^+ l^-$}&\multicolumn{5}{c||}{$M_{T'}=1.0$ TeV search}&\multicolumn{5}{c|}{$M_{T'}=1.5$ TeV search}\\\cline{2-11}
&Signal&$Z+X$&$Z+t$&$S/B$&$S/\sqrt{B}\, (100 \, \mathrm{fb}^{-1})$& Signal &$Z+X$&$Z+t$&$S/B$&$S/\sqrt{B} \,(100 \, \mathrm{fb}^{-1})$\\\hline
Basic cuts                			&1.1             &750       &1.3      &0.0014        &0.39          &0.30          &170              &0.36             &0.0018      &0.23     \\
$Ov^t_{3} > 0.6$        		&0.71             &71        &0.61    &0.010        &0.85             &0.24         &19                &0.14               &0.012    &0.54     \\
$b$-tag                    			&0.52           &1.6       &0.42     &0.25         &3.6                &0.15         &0.36             &0.086            &0.33       &2.2   \\
$\Delta R_{ll}<1.0$	                 &0.52           &1.6       &0.41      &0.26       &3.6               &0.15         &0.36              &0.086          &0.33     &2.2      \\
$|m_{ll}-m_Z| < 10$ GeV		&\bf{0.47}     &1.5       &0.37      &\bf{0.26} &\bf{3.5}     &\bf{0.13}  &0.33               &0.078           &\bf{0.32}      &\bf{2.1}\\\hline
$N_\mathrm{fwd}\geq 1$      	&\bf{0.29}     &0.23     &0.11      &\bf{0.88}   &\bf{5.1}     &\bf{0.088}   &0.051            &0.019           &\bf{1.3}      &\bf{3.3}\\\hline
\end{tabular}
\caption{Example-cutflow for signal- and background events in the $Z_{\mathrm{inv}} +t
+j$ search (top) and in the $Z_{ll}+t+j$ channel (bottom) for $\sqrt{s} = 14 \, \mbox{TeV}$. Cross sections after the respective cuts for signal and backgrounds are given in fb. The $S/\sqrt{B}$ values are given for a luminosity of $100\, \mathrm{fb}^{-1}$. The example signals $\sigma_{T'}\equiv\sigma( p p \rightarrow T'/\bar{T}' + X)\times BR(T'\rightarrow tZ)$  displayed here are $80 \, \mbox{fb}$ for $M_T' = 1.0$ TeV searches and $ 24 \, \mbox{fb}$ for $M_T' =1.5$ TeV searches. The corresponding parameter points of our sample model are given in the text.}
\label{tab:cutflow}
\end{table*}
\end{center}

On the parton level, we generate the events with simple generation level cuts on leptons ( $p_T > 10$ GeV, $|\eta| <2.5$, $\Delta R_{ll} > 0.1~(0.4)$) for the $Z_{ll}$ ($Z_{\mathrm{inv}}$) channel and quarks ($p_T > 15$ GeV, $|\eta| <5$, $\Delta R_{jj} > 0.1$). 
We then cluster the showered events using the \verb|FASTJET| \cite{Cacciari:2011ma} implementation of the anti-kT algorithm \cite{Cacciari:2008gp}, where we use $R = 1.0$ for ``fat jets'', $r = 0.4$ for the light and $b$-jets and $r=0.2$ for forward jets.

The main backgrounds for the $\slashed{E}_T$ channel are SM processes containing a $Z$ boson in the final state, as well as the SM $t\bar{t}$ production.  The ``$Z$-containing'' backgrounds include $Z + t$, characterized by a true $Z$ and a top quark, where we include $Z t\bar{t}$ and $Z t/\bar{t}$ (with up to two extra jets) into our simulation.
Similarly, we define $Z+X$ background to be SM events which contain a $Z$ and ``fake'' (hadronic) top signal. In this class, we include $Z$, $Z b\bar{b}$, $Z+Z/W$ with up to  two additional jets. 
Finally, we include  $t\bar{t}$ background with up to two additional jets.

Di-lepton channels are afflicted by similar SM backgrounds, with the exception of $t\bar{t}$ which is effectively vetoed by requiring two hard leptons which reconstruct a $Z$ mass, small missing energy. 
\bigskip

\begin{center}
\textbf{III. $T'$ SEARCH STRATEGY FOR LHC RUN II}
\end{center}

We choose the cut schemes for the $T'$ search in $t+ j + Z_{\mathrm{inv}}$ and the $t + j + Z_{l\bar{l}}$ channels to be identical for cuts focused on the $t+j$ part of the event in order to allow for a fair comparison.    As a part of the ``basic cuts'', we demand at least one fat jet  with $p_T^{\rm fj}> 400$ GeV (600 GeV) for the $M_{T'}$=1 TeV (1.5 TeV) searches with  $|\eta_{\rm fj}| <2.5$ as well as $p_T^j > 25$ GeV for light, $b$, and forward jets. For the $Z_{\mathrm{inv}}$ channel, we furthermore require  absence of any isolated leptons (mini-ISO $> 0.7$ \cite{Rehermann:2010vq}) with $p^l_T > 25$ GeV while for the $Z_{ll}$ channel we instead follow a modified prescription of Ref.~\cite{Reuter:2014iya}, requiring at least two isolated leptons with $p_T^l > 25$ GeV. The two hardest leptons are then required to reconstruct a leptonic $Z$ boson candidate. We demand $p^z_T > 225$ GeV and $|\eta_z| < 2.3$. Finally, for the $Z_{ll}$-channel, we demand $\Delta R_{ll} < 1.0$ and $|m_{ll}-m_Z| < 10$ GeV.

For top identification  we follow a procedure analogous to Ref. \cite{Backovic:2014uma}, based on the \verb|TEMPLATE TAGGER v.1.0| \cite{Backovic:2012jk} implementation of the template overlap method \cite{Almeida:2011aa, Almeida:2010pa, Backovic:2012jj, Backovic:2013bga}. For a $R=1.0$ jet to be tagged as a ``top,'' we demand a 3 body top template overlap score of $Ov^t_{3}>0.6$. 

We require every fat jet which passes the top selection criteria to also be $b$-tagged, whereby we define a ``fat jet $b$-tag'' as presence of at least one $b$-tagged $r=0.4$ jet within the fat jet ($\Delta R_{tb} < 1.0)$. We assume $b$-tagging efficiencies of $75 \%$ for every $b$ jet to be tagged as a $b$, with a fake rate of $18 \%$ and $1\%$ for $c$ and light jets respectively.

We further utilize the fact that the spectator light jet in the signal events is typically emitted at low $|\eta|$, a very special feature of singly produced $T'$ event topology. For the purpose of ``forward jet tagging'' we  recluster the events with a cone radius $r=0.2$ and demand  at least one forward jet with $p^{\rm fwd}_T > 25$ GeV and $2.5 < \eta^{\rm fwd} < 4.5$ (see Refs.\cite{Backovic:2014uma, DeSimone:2012fs} for a discussion and evaluation of this forward-jet-tagging procedure).

 In the $Z_{\mathrm{inv}}$ channel, we impose a strong cut of $\slashed{E}_T > 400$ GeV (600 GeV) for the $M_{T'}$=1 TeV (1.5 TeV) search. The $t\bar{t}$ background of the invisible $Z$ channel can be further suppressed by demanding the reconstructed  $\slashed{E}_T$ to be isolated from any hard jets by  $|\Delta \phi_{\slashed{E}_T j}| > 1$. 
 

 \begin{center}
 \begin{figure*}[!]
 \begin{tabular}{cc}
 \includegraphics[width=.4\textwidth]{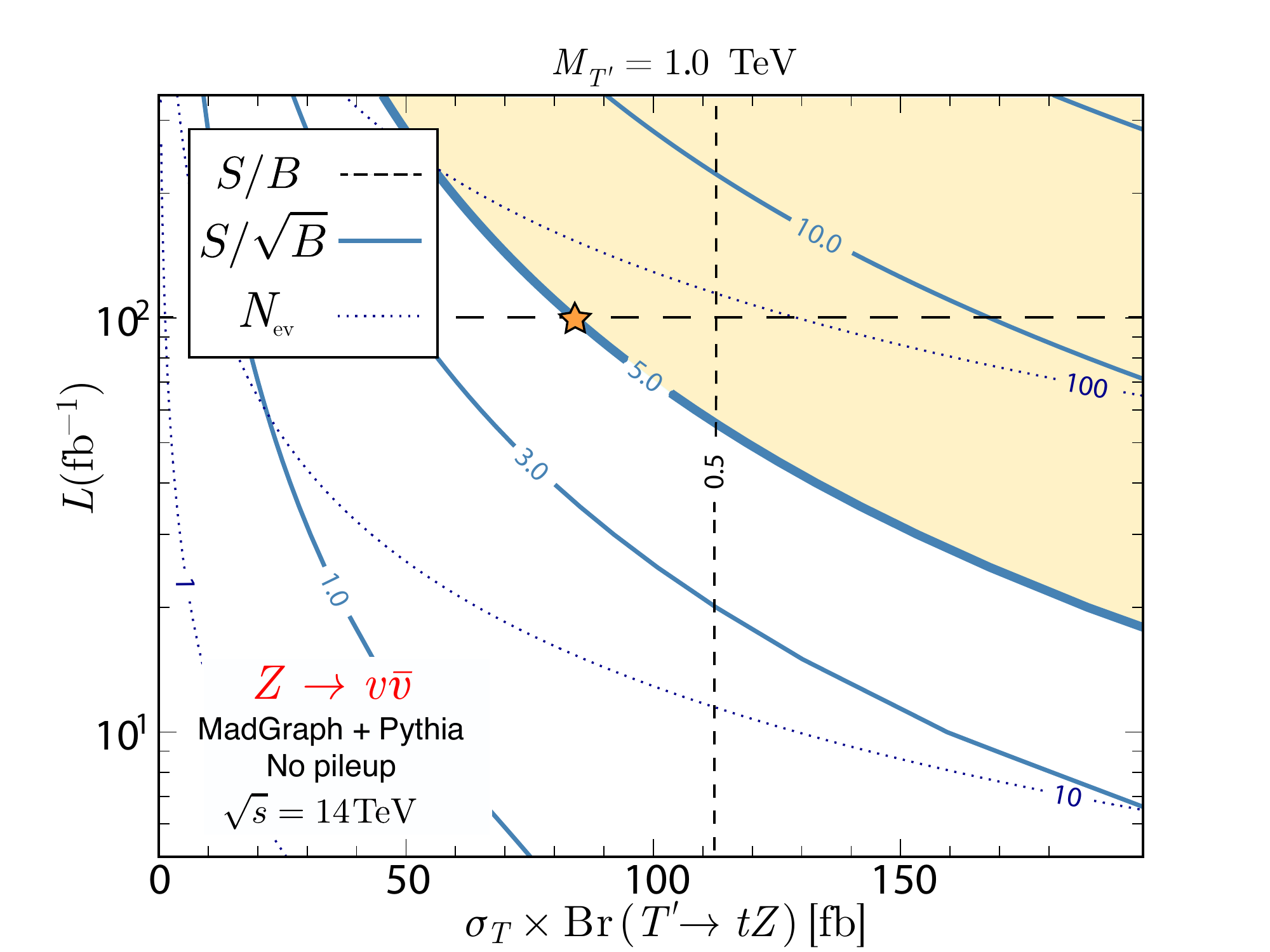}
 \includegraphics[width=.4\textwidth]{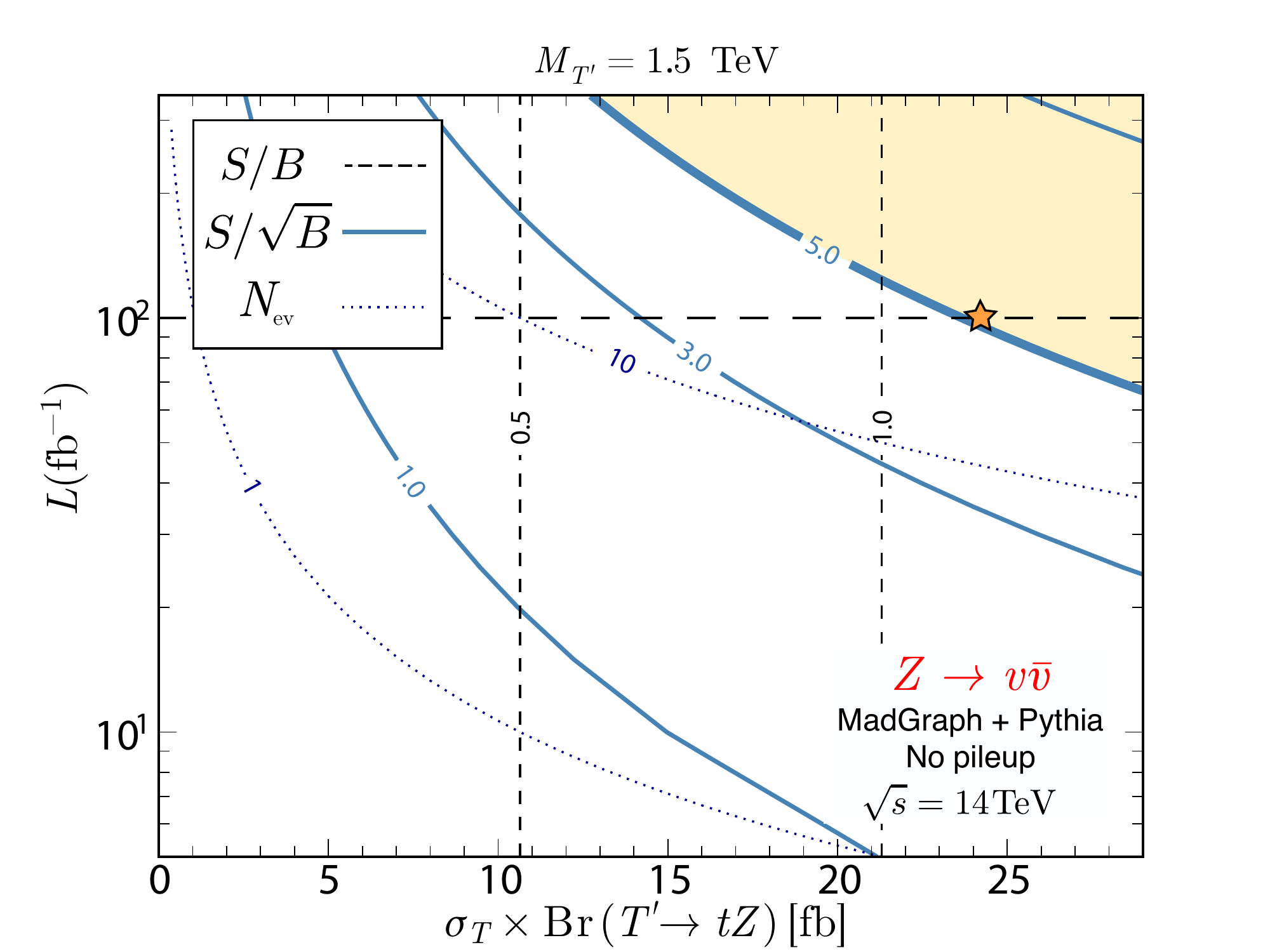}\\
  \includegraphics[width=.4\textwidth]{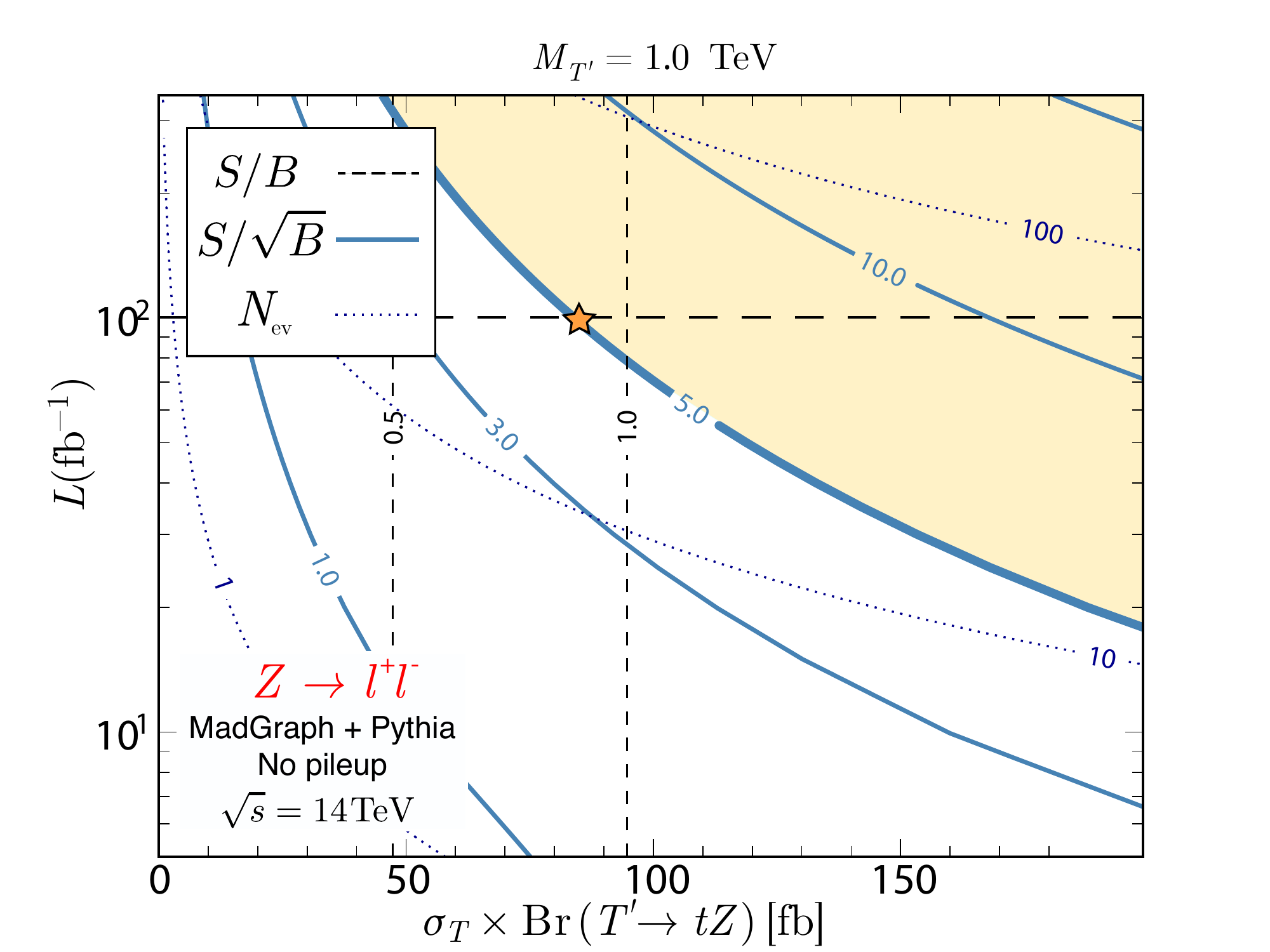}
 \includegraphics[width=.4\textwidth]{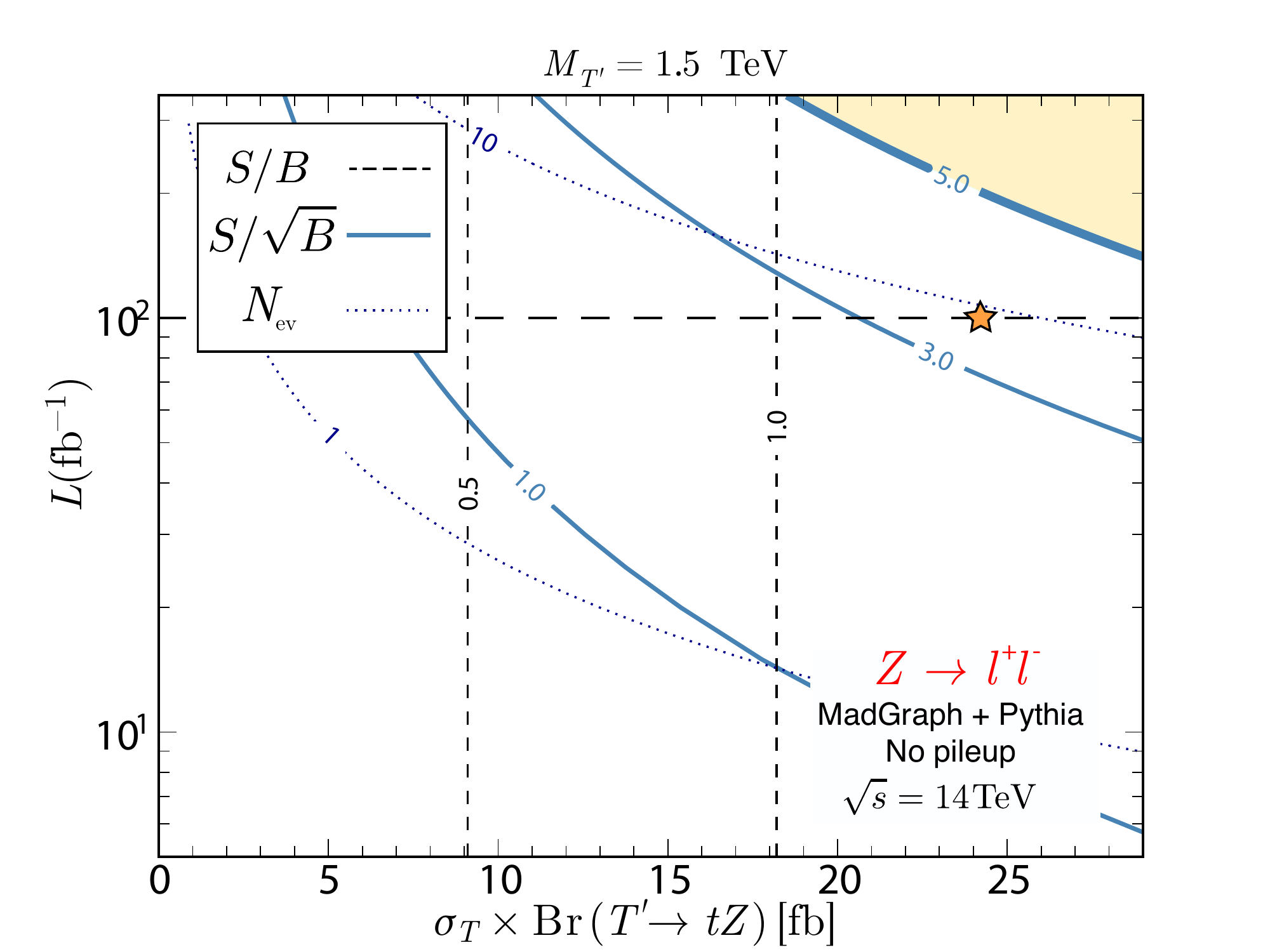}
 \end{tabular}
 \caption{Reach of Run II for discovering $T'$ states of $M_{T'} =1$ TeV (left) and $M_{T'}=1.5$ TeV (right). Shaded regions, signifying $S/\sqrt{B} > 5$ and number of signal events $N_{\mathrm{ev}} > 10$, illustrate the discovery reach of LHC Run II. The reference model points from Table \ref{tab:cutflow} are marked with a star. The horizontal dashed line in each plot marks the benchmark integrated luminosity of $100 \,\mathrm{fb}^{-1}$. The reference production cross sections on the $x$ axis assume a lower cut of $p_T^{j} > 15$ GeV, for the spectator quarks.}
 \label{fig:resultsmet}
 \end{figure*}
 \end{center}
 
 \begin{center}
\textbf{IV. RESULTS}
\end{center}

  Table \ref{tab:cutflow} shows an example cut flow. The sample signal cross sections given correspond to the parameter point $(f =780 \mbox{ GeV} , M_1 =810 \mbox{ GeV}, \lambda_L =2.1, \lambda_R =0.67 $) for the 1 TeV partner search and  $(f = 780 \mbox{ GeV}, M_1 =1.3 \mbox{ TeV} , \lambda_L =3, \lambda_R =0.7$) for the 1.5 TeV partner search. These parameter points yield a large production cross section within the model defined by Eq.(\ref{eq:Leff}),  representing  optimistic scenarios in the searches for $T'$ channels.

Our analysis shows that jet substructure techniques ($Ov^t_{3}$) in conjunction with fat jet $b$-tagging efficiently diminish top quark free background channels (e.g. $Z+X$). 
For the $Z_{\mathrm{inv}}$ search, the $\slashed{E}_T$ cut becomes a crucial discriminant, in particular for $t\bar{t}$ background, with the factor of $\sim 10$ ($\sim 20$) improvement in signal to background ratio ($S/B$) in the 1~TeV (1.5~TeV) search. Forward jet tagging provides a significant improvement in both, $Z_{\mathrm{inv}}$ and $Z_{ll}$ channels, where we find a factor of $\sim 4$ improvement in  $S/B$. For $M_{T'} = 1 $ TeV, the performance of the $Z_{\mathrm{inv}}$ channel is comparable to $Z_{ll},$ in terms of signal significance ($S/\sqrt{B}$), with the former yielding a larger signal cross section, but the latter giving a better $S/B$. If we go to higher masses, such as $M_{T'} = 1.5$ TeV, Table \ref{tab:cutflow} already suggests that the $Z_{\mathrm{inv}}$ channel outperforms di-leptons.

Figure  \ref{fig:resultsmet} illustrates the results in a more complete fashion. Results for the $Z_{\mathrm{inv}}$ and $Z_{ll}$ searches are given in the top and bottom panel for $M_{T'} = 1$ TeV (left) and $M_{T'} = 1.5$ TeV (right). In each plot,  the solid contours show  $S/\sqrt{B}$, while the dotted lines show the number of signal events after cuts, as a function of luminosity and $\sigma_{T'}$. The shaded areas show regions where we expect $S/\sqrt{B} \geq 5$ and at least ten signal events. 
Both the $Z_{\mathrm{inv}}$ and the $Z_{ll}$ channels yield comparable discovery potential for top partners with masses of $\sim 1$ TeV, assuming efficient forward jet tagging from the previous section. For a luminosity of 100 fb$^{-1}$ (dashed horizontal line), optimistic cross sections of $\sigma_{T'} \gtrsim 80 $ fb can be probed at $S/\sqrt{B} \geq 5$ with $N > 10$ signal events.  A comparison of shaded areas in the left panels of Fig. \ref{fig:resultsmet} shows that the missing energy channel becomes important at $M_{T'} \sim 1$ TeV, where inclusion of the $Z_{\mathrm{inv}}$ channel in this mass range would greatly complement the di-lepton searches. 

Probing masses higher than 1 TeV yields a different scenario, as shown in the right panels of Fig. \ref{fig:resultsmet} for the case of  $M_{T'} = 1.5$ TeV.  A partner of mass $M_{T'} = 1.5$   and a cross section $\sigma_{T'} = 24\, \mathrm{fb}$ can be discovered in the $Z_{\mathrm{inv}}$ channel with $100\, \mathrm{fb}^{-1}, $ whereas  approximately $200\, \mathrm{fb}^{-1}$ would be required in the $Z_{ll}$ channel to become sensitive to the same cross section. Considerations of lower signal cross sections yield the same conclusion, as the $Z_{\mathrm{inv}}$ channel gives both better sensitivity and a higher number of signal events at a fixed luminosity over the entire space of reasonable signal cross sections. 
\bigskip   

\begin{center}
\textbf{V. CONCLUSIONS} 
\end{center}

We discussed the LHC Run II  potential to discover  new heavy physics in the boosted $Z$ + fat jet channel.
As an illustration,  we focused on $t + j+ \slashed{E}_T $ and the $t + j+ $~di-lepton search channels for a vectorlike top partner $T'$ decaying into $tZ$ which occurs in a large class of SM extensions. 

Our main conclusion is that future considerations of  the $Z_{\mathrm{inv}}$ channel in searches for new physics in the $Z + t,h,W$ channel will greatly extend the ability of the early LHC Run II to discover possible new particles at the TeV scale. The di-lepton channel maintains competitive sensitivity for $M_{T'} \approx 1$ TeV, with a higher $S/B$ but lower signal cross section. 
The situation changes rapidly above 1 TeV. At $M_{T'} \approx 1.5$ TeV, we show that the $Z_{\mathrm{inv}}$ channel  displays clear superior performance in prospects for discovering the $T'$ states, with both larger signal significance and number of events. The di-lepton channel still remains important, as the event reconstruction capability using leptons has an advantage over large missing energy. The discovery reach can be further improved by combining the two channels.  Furthermore, we show that the discovery potential can be substantially improved by demanding a high-energy forward-jet tagging in both $Z_{\mathrm{inv}}$ and $Z_{ll}$ channels.

Our results are a direct consequence of the fact that for $M_T'\gtrsim 1 $ TeV, kinematics of boosted $Z$ decays allow for efficient use of channels with high $\slashed{E_T}$.  A high $\slashed{E}_T$ cut efficiently removes the $t \bar{t}$ background in the $t + \slashed{E}_T$ channel, while forward jet tagging greatly improves the performance in both channels. 

We base our conclusions on an example study of searches for charge 2/3 vectorlike quarks, but the qualitative argument applies more broadly to 
searches for heavy resonances in channels containing a boosted $Z$ boson in the final state. The kinematics of heavy resonance decays in general are mainly determined by the mass of the resonance, while the model dependent details of coupling strength and spin of the resonance only affect the overall production rate. The structure of the vertex which defines the interaction of the heavy resonance with the $Z$ can affect the kinematic distributions (i.e. spin correlations), but these are typically subleading effects. It hence follows that in a fairly generic scenario of heavy resonance decays, there always exists a resonance mass range in which $Z\rightarrow \nu \bar{\nu}$ channel will be more sensitive than $Z\rightarrow l^+ l^-$.

Our analysis suggests that channels with large missing energy will be more sensitive to \textit{discovery} of vectorlike top partners of $\sim$ TeV mass scale at low integrated luminosity. In case a signal is observed, the di-lepton channel will still be important for accurate measurements of the resonance properties, in particular its reconstructed mass and its spin. Due to the superior ability of ATLAS and CMS to reconstruct and measure leptons, compared to missing energy reconstruction, it is likely that the di-lepton channel will provide better measurements of the $T^\prime$ mass. However, past experience has shown that methods employing kinematic edges, such as $m_{T2}$ \cite{Lester:1999tx}, could result in competitive mass measurements in channels involving large missing energy. As here we were interested mostly in prospects for discovery of new physics, we chose to postpone the comparative analysis of mass measurements for future analyses.

Effects of pileup contamination should be considered in future analyses, especially considering the high instantaneous luminosity expected for Run II. However, works of Refs.~\cite{Backovic:2013bga,Backovic:2014uma} have shown that effects of pileup on $\slashed{E_T}$, $Ov_3^t$, forward jet tagging and $b$-tagging can be effectively mitigated, even at 50 interactions per bunch crossing, without requiring exotic pileup subtraction techniques. We hence expect our conclusions to be robust even when considering high pileup levels. 

\bigskip 
\begin{center}  
\textbf{ACKNOWLEDGMENTS}
\end{center}

 The authors would like to thank the Weizmann theory group and the organizers of the Naturalness 2014 workshop for the hospitality during the initial stages of this project. This work was supported by the National Research Foundation of Korea(NRF) grant funded by the Korea government(MEST) (No. 2012R1A2A2A01045722), and Basic Science Research Program through the National Research Foundation of Korea(NRF) funded by the ministry of Education, Science and Technology (No. 2013R1A1A1062597). This work is also supported by HTCaaS group of KISTI (Korea Institute of Science and Technology Information). S. L. and T. F. are also supported by Korea-ERC researcher visiting program through the National Research Foundation of Korea(NRF) (No. 2014K2a7B044399 and No. 2014K2a7A1044408). M. B. is in part supported by the Belgian Federal Science Policy Office through the Inter-university Attraction Pole P7/37.

\bibliography{draft}

\begin{thebibliography}{27}
\expandafter\ifx\csname natexlab\endcsname\relax\def\natexlab#1{#1}\fi
\expandafter\ifx\csname bibnamefont\endcsname\relax
  \def\bibnamefont#1{#1}\fi
\expandafter\ifx\csname bibfnamefont\endcsname\relax
  \def\bibfnamefont#1{#1}\fi
\expandafter\ifx\csname citenamefont\endcsname\relax
  \def\citenamefont#1{#1}\fi
\expandafter\ifx\csname url\endcsname\relax
  \def\url#1{\texttt{#1}}\fi
\expandafter\ifx\csname urlprefix\endcsname\relax\def\urlprefix{URL }\fi
\providecommand{\bibinfo}[2]{#2}
\providecommand{\eprint}[2][]{\url{#2}}

\bibitem[{\citenamefont{Schmaltz and Tucker-Smith}(2005)}]{Schmaltz:2005ky}
\bibinfo{author}{\bibfnamefont{M.}~\bibnamefont{Schmaltz}} \bibnamefont{and}
  \bibinfo{author}{\bibfnamefont{D.}~\bibnamefont{Tucker-Smith}},
  \bibinfo{journal}{Ann.Rev.Nucl.Part.Sci.} \textbf{\bibinfo{volume}{55}},
  \bibinfo{pages}{229} (\bibinfo{year}{2005}), \eprint{hep-ph/0502182}.

\bibitem[{\citenamefont{Contino et~al.}(2003)\citenamefont{Contino, Nomura, and
  Pomarol}}]{Contino:2003ve}
\bibinfo{author}{\bibfnamefont{R.}~\bibnamefont{Contino}},
  \bibinfo{author}{\bibfnamefont{Y.}~\bibnamefont{Nomura}}, \bibnamefont{and}
  \bibinfo{author}{\bibfnamefont{A.}~\bibnamefont{Pomarol}},
  \bibinfo{journal}{Nucl.Phys.} \textbf{\bibinfo{volume}{B671}},
  \bibinfo{pages}{148} (\bibinfo{year}{2003}), \eprint{hep-ph/0306259}.

\bibitem[{\citenamefont{Agashe et~al.}(2005)\citenamefont{Agashe, Contino, and
  Pomarol}}]{Agashe:2004rs}
\bibinfo{author}{\bibfnamefont{K.}~\bibnamefont{Agashe}},
  \bibinfo{author}{\bibfnamefont{R.}~\bibnamefont{Contino}}, \bibnamefont{and}
  \bibinfo{author}{\bibfnamefont{A.}~\bibnamefont{Pomarol}},
  \bibinfo{journal}{Nucl.Phys.} \textbf{\bibinfo{volume}{B719}},
  \bibinfo{pages}{165} (\bibinfo{year}{2005}), \eprint{hep-ph/0412089}.

\bibitem[{\citenamefont{Agashe et~al.}(2006)\citenamefont{Agashe, Contino,
  Da~Rold, and Pomarol}}]{Agashe:2006at}
\bibinfo{author}{\bibfnamefont{K.}~\bibnamefont{Agashe}},
  \bibinfo{author}{\bibfnamefont{R.}~\bibnamefont{Contino}},
  \bibinfo{author}{\bibfnamefont{L.}~\bibnamefont{Da~Rold}}, \bibnamefont{and}
  \bibinfo{author}{\bibfnamefont{A.}~\bibnamefont{Pomarol}},
  \bibinfo{journal}{Phys.Lett.} \textbf{\bibinfo{volume}{B641}},
  \bibinfo{pages}{62} (\bibinfo{year}{2006}), \eprint{hep-ph/0605341}.

\bibitem[{\citenamefont{Aad et~al.}(2014)}]{Aad:2014efa}
\bibinfo{author}{\bibfnamefont{G.}~\bibnamefont{Aad}} \bibnamefont{et~al.}
  (\bibinfo{collaboration}{ATLAS Collaboration}), \bibinfo{journal}{JHEP}
  \textbf{\bibinfo{volume}{1411}}, \bibinfo{pages}{104} (\bibinfo{year}{2014}),
  \eprint{1409.5500}.

\bibitem[{\citenamefont{Reuter and Tonini}(2014)}]{Reuter:2014iya}
\bibinfo{author}{\bibfnamefont{J.}~\bibnamefont{Reuter}} \bibnamefont{and}
  \bibinfo{author}{\bibfnamefont{M.}~\bibnamefont{Tonini}}
  (\bibinfo{year}{2014}), \eprint{1409.6962}.

\bibitem[{\citenamefont{Butterworth et~al.}(2008)\citenamefont{Butterworth,
  Davison, Rubin, and Salam}}]{Butterworth:2008iy}
\bibinfo{author}{\bibfnamefont{J.~M.} \bibnamefont{Butterworth}},
  \bibinfo{author}{\bibfnamefont{A.~R.} \bibnamefont{Davison}},
  \bibinfo{author}{\bibfnamefont{M.}~\bibnamefont{Rubin}}, \bibnamefont{and}
  \bibinfo{author}{\bibfnamefont{G.~P.} \bibnamefont{Salam}},
  \bibinfo{journal}{Phys.Rev.Lett.} \textbf{\bibinfo{volume}{100}},
  \bibinfo{pages}{242001} (\bibinfo{year}{2008}), \eprint{0802.2470}.

\bibitem[{\citenamefont{Chen et~al.}(2014)\citenamefont{Chen, Davoudiasl, and
  Kim}}]{Chen:2014oha}
\bibinfo{author}{\bibfnamefont{C.-Y.} \bibnamefont{Chen}},
  \bibinfo{author}{\bibfnamefont{H.}~\bibnamefont{Davoudiasl}},
  \bibnamefont{and} \bibinfo{author}{\bibfnamefont{D.}~\bibnamefont{Kim}},
  \bibinfo{journal}{Phys.Rev.} \textbf{\bibinfo{volume}{D89}},
  \bibinfo{pages}{096007} (\bibinfo{year}{2014}), \eprint{1403.3399}.

\bibitem[{\citenamefont{Aad et~al.}()}]{ATLASnotes1}
\bibinfo{author}{\bibfnamefont{G.}~\bibnamefont{Aad}} \bibnamefont{et~al.}
  (\bibinfo{collaboration}{ATLAS Collaboration}),
  \bibinfo{note}{{ATLAS-CONF-2014-036, ATLAS-COM-CONF-2014-055}}.

\bibitem[{\citenamefont{Chatrchyan et~al.}(2014)}]{Chatrchyan:2013uxa}
\bibinfo{author}{\bibfnamefont{S.}~\bibnamefont{Chatrchyan}}
  \bibnamefont{et~al.} (\bibinfo{collaboration}{CMS Collaboration}),
  \bibinfo{journal}{Phys.Lett.} \textbf{\bibinfo{volume}{B729}},
  \bibinfo{pages}{149} (\bibinfo{year}{2014}), \eprint{1311.7667}.

\bibitem[{\citenamefont{Backovic
  et~al.}(2014{\natexlab{a}})\citenamefont{Backovic, Perez, Flacke, and
  Lee}}]{Backovic:2014uma}
\bibinfo{author}{\bibfnamefont{M.}~\bibnamefont{Backovic}},
  \bibinfo{author}{\bibfnamefont{G.}~\bibnamefont{Perez}},
  \bibinfo{author}{\bibfnamefont{T.}~\bibnamefont{Flacke}}, \bibnamefont{and}
  \bibinfo{author}{\bibfnamefont{S.~J.} \bibnamefont{Lee}}
  (\bibinfo{year}{2014}{\natexlab{a}}), \eprint{1409.0409}.

\bibitem[{\citenamefont{Chatrchyan et~al.}(2012)}]{Chatrchyan:2012ep}
\bibinfo{author}{\bibfnamefont{S.}~\bibnamefont{Chatrchyan}}
  \bibnamefont{et~al.} (\bibinfo{collaboration}{CMS}), \bibinfo{journal}{JHEP}
  \textbf{\bibinfo{volume}{1212}}, \bibinfo{pages}{035} (\bibinfo{year}{2012}),
  \eprint{1209.4533}.

\bibitem[{\citenamefont{Maltoni and Stelzer}(2003)}]{Maltoni:2002qb}
\bibinfo{author}{\bibfnamefont{F.}~\bibnamefont{Maltoni}} \bibnamefont{and}
  \bibinfo{author}{\bibfnamefont{T.}~\bibnamefont{Stelzer}},
  \bibinfo{journal}{JHEP} \textbf{\bibinfo{volume}{0302}}, \bibinfo{pages}{027}
  (\bibinfo{year}{2003}), \eprint{hep-ph/0208156}.

\bibitem[{\citenamefont{Ball et~al.}(2013)}]{Ball:2013gsa}
\bibinfo{author}{\bibfnamefont{R.~D.} \bibnamefont{Ball}} \bibnamefont{et~al.}
  (\bibinfo{collaboration}{The NNPDF Collaboration}),
  \bibinfo{journal}{Phys.Lett.} \textbf{\bibinfo{volume}{B723}},
  \bibinfo{pages}{330} (\bibinfo{year}{2013}), \eprint{1303.1189}.

\bibitem[{\citenamefont{Sjostrand et~al.}(2006)\citenamefont{Sjostrand, Mrenna,
  and Skands}}]{Sjostrand:2006za}
\bibinfo{author}{\bibfnamefont{T.}~\bibnamefont{Sjostrand}},
  \bibinfo{author}{\bibfnamefont{S.}~\bibnamefont{Mrenna}}, \bibnamefont{and}
  \bibinfo{author}{\bibfnamefont{P.~Z.} \bibnamefont{Skands}},
  \bibinfo{journal}{JHEP} \textbf{\bibinfo{volume}{0605}}, \bibinfo{pages}{026}
  (\bibinfo{year}{2006}), \eprint{hep-ph/0603175}.

\bibitem[{\citenamefont{Mangano et~al.}(2007)\citenamefont{Mangano, Moretti,
  Piccinini, and Treccani}}]{Mangano:2006rw}
\bibinfo{author}{\bibfnamefont{M.~L.} \bibnamefont{Mangano}},
  \bibinfo{author}{\bibfnamefont{M.}~\bibnamefont{Moretti}},
  \bibinfo{author}{\bibfnamefont{F.}~\bibnamefont{Piccinini}},
  \bibnamefont{and} \bibinfo{author}{\bibfnamefont{M.}~\bibnamefont{Treccani}},
  \bibinfo{journal}{JHEP} \textbf{\bibinfo{volume}{0701}}, \bibinfo{pages}{013}
  (\bibinfo{year}{2007}), \eprint{hep-ph/0611129}.

\bibitem[{\citenamefont{Cacciari et~al.}(2012)\citenamefont{Cacciari, Salam,
  and Soyez}}]{Cacciari:2011ma}
\bibinfo{author}{\bibfnamefont{M.}~\bibnamefont{Cacciari}},
  \bibinfo{author}{\bibfnamefont{G.~P.} \bibnamefont{Salam}}, \bibnamefont{and}
  \bibinfo{author}{\bibfnamefont{G.}~\bibnamefont{Soyez}},
  \bibinfo{journal}{Eur.Phys.J.} \textbf{\bibinfo{volume}{C72}},
  \bibinfo{pages}{1896} (\bibinfo{year}{2012}), \eprint{1111.6097}.

\bibitem[{\citenamefont{Cacciari et~al.}(2008)\citenamefont{Cacciari, Salam,
  and Soyez}}]{Cacciari:2008gp}
\bibinfo{author}{\bibfnamefont{M.}~\bibnamefont{Cacciari}},
  \bibinfo{author}{\bibfnamefont{G.~P.} \bibnamefont{Salam}}, \bibnamefont{and}
  \bibinfo{author}{\bibfnamefont{G.}~\bibnamefont{Soyez}},
  \bibinfo{journal}{JHEP} \textbf{\bibinfo{volume}{0804}}, \bibinfo{pages}{063}
  (\bibinfo{year}{2008}), \eprint{0802.1189}.

\bibitem[{\citenamefont{Rehermann and Tweedie}(2011)}]{Rehermann:2010vq}
\bibinfo{author}{\bibfnamefont{K.}~\bibnamefont{Rehermann}} \bibnamefont{and}
  \bibinfo{author}{\bibfnamefont{B.}~\bibnamefont{Tweedie}},
  \bibinfo{journal}{JHEP} \textbf{\bibinfo{volume}{1103}}, \bibinfo{pages}{059}
  (\bibinfo{year}{2011}), \eprint{1007.2221}.

\bibitem[{\citenamefont{Backovic and Juknevich}(2014)}]{Backovic:2012jk}
\bibinfo{author}{\bibfnamefont{M.}~\bibnamefont{Backovic}} \bibnamefont{and}
  \bibinfo{author}{\bibfnamefont{J.}~\bibnamefont{Juknevich}},
  \bibinfo{journal}{Comput.Phys.Commun.} \textbf{\bibinfo{volume}{185}},
  \bibinfo{pages}{1322} (\bibinfo{year}{2014}), \eprint{1212.2978}.

\bibitem[{\citenamefont{Almeida et~al.}(2012)\citenamefont{Almeida, Erdogan,
  Juknevich, Lee, Perez et~al.}}]{Almeida:2011aa}
\bibinfo{author}{\bibfnamefont{L.~G.} \bibnamefont{Almeida}},
  \bibinfo{author}{\bibfnamefont{O.}~\bibnamefont{Erdogan}},
  \bibinfo{author}{\bibfnamefont{J.}~\bibnamefont{Juknevich}},
  \bibinfo{author}{\bibfnamefont{S.~J.} \bibnamefont{Lee}},
  \bibinfo{author}{\bibfnamefont{G.}~\bibnamefont{Perez}},
  \bibnamefont{et~al.}, \bibinfo{journal}{Phys.Rev.}
  \textbf{\bibinfo{volume}{D85}}, \bibinfo{pages}{114046}
  (\bibinfo{year}{2012}), \eprint{1112.1957}.

\bibitem[{\citenamefont{Almeida et~al.}(2010)\citenamefont{Almeida, Lee, Perez,
  Sterman, and Sung}}]{Almeida:2010pa}
\bibinfo{author}{\bibfnamefont{L.~G.} \bibnamefont{Almeida}},
  \bibinfo{author}{\bibfnamefont{S.~J.} \bibnamefont{Lee}},
  \bibinfo{author}{\bibfnamefont{G.}~\bibnamefont{Perez}},
  \bibinfo{author}{\bibfnamefont{G.}~\bibnamefont{Sterman}}, \bibnamefont{and}
  \bibinfo{author}{\bibfnamefont{I.}~\bibnamefont{Sung}},
  \bibinfo{journal}{Phys.Rev.} \textbf{\bibinfo{volume}{D82}},
  \bibinfo{pages}{054034} (\bibinfo{year}{2010}), \eprint{1006.2035}.

\bibitem[{\citenamefont{Backovic et~al.}(2013)\citenamefont{Backovic,
  Juknevich, and Perez}}]{Backovic:2012jj}
\bibinfo{author}{\bibfnamefont{M.}~\bibnamefont{Backovic}},
  \bibinfo{author}{\bibfnamefont{J.}~\bibnamefont{Juknevich}},
  \bibnamefont{and} \bibinfo{author}{\bibfnamefont{G.}~\bibnamefont{Perez}},
  \bibinfo{journal}{JHEP} \textbf{\bibinfo{volume}{1307}}, \bibinfo{pages}{114}
  (\bibinfo{year}{2013}), \eprint{1212.2977}.

\bibitem[{\citenamefont{Backovic
  et~al.}(2014{\natexlab{b}})\citenamefont{Backovic, Gabizon, Juknevich, Perez,
  and Soreq}}]{Backovic:2013bga}
\bibinfo{author}{\bibfnamefont{M.}~\bibnamefont{Backovic}},
  \bibinfo{author}{\bibfnamefont{O.}~\bibnamefont{Gabizon}},
  \bibinfo{author}{\bibfnamefont{J.}~\bibnamefont{Juknevich}},
  \bibinfo{author}{\bibfnamefont{G.}~\bibnamefont{Perez}}, \bibnamefont{and}
  \bibinfo{author}{\bibfnamefont{Y.}~\bibnamefont{Soreq}},
  \bibinfo{journal}{JHEP} \textbf{\bibinfo{volume}{1404}}, \bibinfo{pages}{176}
  (\bibinfo{year}{2014}{\natexlab{b}}), \eprint{1311.2962}.

\bibitem[{\citenamefont{De~Simone et~al.}(2013)\citenamefont{De~Simone,
  Matsedonskyi, Rattazzi, and Wulzer}}]{DeSimone:2012fs}
\bibinfo{author}{\bibfnamefont{A.}~\bibnamefont{De~Simone}},
  \bibinfo{author}{\bibfnamefont{O.}~\bibnamefont{Matsedonskyi}},
  \bibinfo{author}{\bibfnamefont{R.}~\bibnamefont{Rattazzi}}, \bibnamefont{and}
  \bibinfo{author}{\bibfnamefont{A.}~\bibnamefont{Wulzer}},
  \bibinfo{journal}{JHEP} \textbf{\bibinfo{volume}{1304}}, \bibinfo{pages}{004}
  (\bibinfo{year}{2013}), \eprint{1211.5663}.

\bibitem[{\citenamefont{Lester and Summers}(1999)}]{Lester:1999tx}
\bibinfo{author}{\bibfnamefont{C.}~\bibnamefont{Lester}} \bibnamefont{and}
  \bibinfo{author}{\bibfnamefont{D.}~\bibnamefont{Summers}},
  \bibinfo{journal}{Phys.Lett.} \textbf{\bibinfo{volume}{B463}},
  \bibinfo{pages}{99} (\bibinfo{year}{1999}), \eprint{hep-ph/9906349}.

\bibitem[{\citenamefont{Cacciapaglia et~al.}(2015)\citenamefont{Cacciapaglia,
  Cai, Flacke, Lee, Parolini et~al.}}]{Cacciapaglia:2015dsa}
\bibinfo{author}{\bibfnamefont{G.}~\bibnamefont{Cacciapaglia}},
  \bibinfo{author}{\bibfnamefont{H.}~\bibnamefont{Cai}},
  \bibinfo{author}{\bibfnamefont{T.}~\bibnamefont{Flacke}},
  \bibinfo{author}{\bibfnamefont{S.~J.} \bibnamefont{Lee}},
  \bibinfo{author}{\bibfnamefont{A.}~\bibnamefont{Parolini}},
  \bibnamefont{et~al.}, \bibinfo{journal}{JHEP}
  \textbf{\bibinfo{volume}{1506}}, \bibinfo{pages}{085} (\bibinfo{year}{2015}),
  \eprint{1501.03818}.

\end{thebibliography}

\end{document}